

\input phyzzx

\tolerance=10000


\Pubnum = {QMW-PH-93-20}
\date = {April 1994}
 \pubtype={}

\titlepage
\title{The Quantum Mechanics of a \lq\lq Spinor-twin" Type II Superparticle.}
\author{Jos\'e-Luis V\'azquez-Bello\footnote\star{On leave from QMW College;
                        BELLO at 138.37.48.201, BELLO at 152.84.253.2 or
                        JLBELLO at 148.247.8.10}
                                                       }
\address{CBPF-CNPq Centro Brasileiro de Pesquisas Fisicas, \break
           Rua Dr. Xavier Sigaud 150, CEP. 22290,\break
           Rio de Janeiro - RJ, BRASIL.}
\address{Physics Department, Queen Mary and Westfield College,\break
             Mile End Road, London E1 4NS,\break
                 UNITED KINGDOM.}

\abstract

{ Ten dimensional supersymmetric Yang-Mills theory may be described,
  in the light-cone gauge, in terms of either a vector or spinor
  superfield satisfying certain projection conditions (type I or II).
  These have been presented in a $ SO(9,1) $ form, and used to
  construct spinning superparticle theories in extended spaces.
  This letter presents the covariant quantisation of a "spinor-twin"
  type II superparticle theory by using the standard techniques
  of Batalin and Vilkovisky. The quantum action defines a quadratic
  field theory, whose ghost-independent BRST cohomology class gives
  the spectrum of N=1 super Yang-Mills. }

\endpage

\pagenumber=1
%

\def\psh {\rlap{/}{p}}

\def\Lamdsh {\Lambda\llap{/}}
\def\lmdash {\Lambda\llap{/}}
\def\Sigmash {\Sigma\llap{/}}
\def\Sigmsh {\Sigma\llap{/}}

\def\xmu { x^\mu}
\def\pmu { p_\mu}
\def\half {{ 1\over 2}}
\def\thetaA {\theta_A}
\def\thetaB {\theta_B}

\def\phiA {\phi^A}
\def\PhiA {\Phi^A}
\def\adot {\dot a}

\def\gammu {{\gamma^\mu}}
\def\gamnu {{\gamma^\nu}}
\def\thetaB {\theta_B}
\def\PsiA {\Psi_A}

\def\grad {\partial}
\def\Sigmnrs {\Sigma_{\mu\nu\rho\sigma}}
\def\Gammu {\Gamma^\mu}

\def\Lamnrs {\Lambda_{\mu\nu\rho\sigma}}
\def\starxmu {x^\star_\mu}
\def\startheta {\theta^\star}
\def\stard {d^\star}
\def\stare {e^\star}
\def\spsi {\psi^\star}
\def\sphi {\phi^\star}
\def\shatphi {{\hat\phi}^\star}
\def\svarphi {\varphi^\star}
\def\sLamdsh {{\Lamdsh}^\star}
\def\sbeta {\beta^\star}
\def\szeta {\zeta^\star}
\def\seta {\eta^\star}
\def\somega {\omega^\star}
\def\skappa {\kappa^\star}
\def\starv {v^\star}
\def\sSigmash {{\Sigmash}^\star}
\def\starc { c^\star}

Discussions of the mechanics of particles with spin shows that
these can be described by either a particle theory with local world-line
supersymmetry
\REF\one{L. Brink, S. Deser, B. Zumino, P. DiVecchia and P.S. Howe,
Phys. Lett. {\bf B64} (1976) 435; F.A. Berezin and M.S. Marinov,
Ann. Phys. {\bf 104} (1977) 336; L. Brink, P. DiVecchia and P.S. Howe,
Nucl. Phys. {\bf B118} (1977) 76; C. Galvao and T. Teitelboim,
J. Math. Phys. {\bf 21} (1980) 1863; P.S. Howe, S. Penati; M. Pernici and
P.K. Townsend, Phys. Lett. {\bf 215} (1988) 555;
V.D. Gershum and V.I. Tkach, JETP Lett. {\bf 29} (1979) 320;
W. Siegel, Int. J. Mod. Phys. {\bf A3} (1988) 2713.}[\one], or
by a local fermionic symmetry\REF\two{C.M.Hull and J.L. V\'azquez-Bello,
{\it Particules, Superparticles and Super-Yand-Mills},
Preprint QMW-PH-93-07, hepth/9308022.}[\two].
This was generalised to superpace, to obtain a number of
{\it spinning} superparticle theories satisfying certain constraints
whose spectrum were precisely those of the ten-dimensional supersymmetric
Yang-Mills theory.
The quantum mechanics of a free superparticle in a ten-dimensional
space-time is of interest because of its close relationship to
ten-dimensional super Yang-Mills theory, and
this corresponds to the massless sector of type I superstring theory.
%
%
The $SO(9,1)$ covariant superfield formulation of super Yang-Mills
which reduces to $SO(8)$ ones may be obtained by either an $SO(9,1)$
vector or spinor superfields. These superfields are chosen to satisfy either
rotational quadratic ( \lq\lq Type I")
or linear (\lq\lq Type II") constraints that restricts their field content
to the physical propagating fields.
The constraints are imposed by an explicit projection operator, constructed
out of super-covariant derivatives, acting on unconstrained superfields.
These were explicitly given on its $SO(8)$ form in
\REF\brink{L. Brink, M.B. Green and J.H. Schwarz, Nucl. Phys.
{\bf B223} (1983) 125.}[\brink],
and presented on its $SO(9,1)$ form in [\two].
The spinor and vector superfields are related by
$\gamma^i_{\adot b}\Psi_b =(1/8)D_{\adot}\Psi^i$.
Yet, it is well known the abstruse quantisation of superparticle
models in a covariant manner
\REF\prob{W. Siegel, Phys. Lett. {\bf B203} (1988) 79;
L. Brink, in {\it Physics and Mathematics of Strings}, eds. L. Brink, D.
Friedan and A.M. Polyakov, (World Scientific, 1990);
A. Mikovi\'c, M.Ro\v cek, W. Siegel, A.E. van de Ven, P. van Nieuwenhuizen
and J.P. Yamron, Phys. Lett. {\bf B235} (1990) 106;
F. E$\beta$ler, E. Laenen, W. Siegel and J.P. Yamron, Phys. Lett. {\bf B254}
(1991) 411; F. E$\beta$ler, M. Hatsuda, E. Laenen, W. Siegel, J.P. Yamron,
T. Kimura and A. Mokovi\'c, Nucl. Phys. {\bf B364} (1991) 67;
E.A. Bergshoeff and R.E. Kallosh, Phys. Lett. {\bf B240} (1990) 105;
Nucl. Phys. {\bf B333} (1990) 605; E.A. Berghsoeff, R.E. Kallosh and
A. van Proeyen, Phys. Lett. {\bf B251} (1990) 128;R.E. Kallosh,
Phys. Lett. {\bf B251} (1990) 134; M. Huq, Int. J. Mod. Phys. {\bf A7}
(1992) 4053; J.L. V\'azquez-Bello, Int. J. of Mod. Phys.
{\bf A19} (1992) 4583.}[\prob],
there are
by now several formulations which can be covariantly quantized.
These superparticle theories with spectra coinciding with
that of the super Yang-Mills are constructed by adding appropriate
Lagrange multiplier terms to certain superparticle actions,
some of them leading to these type I (II) constraints.

This letter presents the covariant quantisation of a \lq\lq spinor-twin"
type II
superparticle model, by using the Batalin-Vilkovisky formalism
\REF\bv{I.A. Batalin and G.A. Vilkovisky, Phys. Lett. {\bf B102}
 (1981) 462;Phys. Rev. {\bf D28} (1983) 2567.}[\bv].
The methods of
\REF\twox{M.B. Green and C.M. Hull, Nucl. Phys. {\bf B344} (1990) 115.}
\REF\twomike{M.B. Green and C.M. Hull, Mod. Phys. Lett. A18 (1990)
1399.}[\twox ,\twomike]
are used to argue that the zero ghost-number BRST cohomology
class in the reduced formalism is exactly the same as the zero ghost-number
cohomology class in the full formalism with an infinite number of ghosts.

We begin by briefly reviewing the description of ten-dimensional
\lq\lq spinor-twin" type I superparticle models [\two].
A spinor wavefunction
can be obtained either from a spinning particle with local world-line
supersymmetry, or from a particle action with local fermionic symmetry.
In [\two], it was seen that super Yang-Mills theory in ten-dimensions
is described by precisely such wavefunctions subject to certain extra
super-covariant constraints.
The quantum mechanics of the spinor-twin type I superparticle theory was
given in [\twomike]. This superparticle action is formulated in an
extended ten-dimensional
superspace with coordinates $(\xmu ,\thetaA ,\phiA )$ where
$\thetaA$ and $\phiA$ are anti-commuting Majorana-Weyl spinors.
\footnote
\spadesuit
{A Majorana spinor $\Psi$ corresponds to a pair of Majorana-Weyl
spinors, $\PsiA$ and $\Psi^A$. The $32 \times 32$ matrices $C\gammu$
(where $C$ is the charge conjugation matrix) are block diagonal with
$16 \times 16$ blocks $\gammu^{AB}$, $\gammu_{AB}$ which are symmetric
and satisfy
$\gammu^{AB}\gamnu_{BC} + \gamnu^{AB}\gammu_{BC} =2\eta^{\mu\nu}\delta^A_C$.
In this notation the supercoordinates has components $\thetaA$,
$\theta\gammu\dot\theta = \thetaA \gammu_{AB} \dot\thetaB$,
$\psh_{AB} = p^\mu \gammu_{AB}$, etc. }
The physical states are described by a superspace wavefunction
satisfying [\two]
$$\eqalign{
p^2 \Psi_A &=0, \qquad \psh_{AC}D^C \Psi_B =0, \qquad \psh^{AB}\Psi_B =0, \cr
&D^AD^B\Psi_C + 8 (\gamma^\mu)^{E[A}(\Gamma_\mu\psh )^{B]}_{\ \ C}\Psi^D =0.
\cr}\eqn\quadra$$
which leaves a superfield $\Psi_a (x^i,\theta^{\adot} )$
satisfying a quadratic projection condition
which is precisely the $SO(8)$ constraint of ten dimensional
super Yang-Mills theory.
The covariant quantisation  of this superparticle was briefly
discussed in [\twomike] in the gauge $e=1$ with the other gauge
fields set to zero. Covariant quantisation requires the methods of
Batalin and Vilkovisky [\bv] since the gauge algebra only
closes on shell, and requires an infinite number of ghost fields
since the symmetries are infinitely reducible.
Following the BV procedure leads to a gauge-fixed quantum action
which, after field redefinitions and integrating out all non-propagating
fields, takes the form [\twomike]
$$
S_Q =\int\! d\tau \Big[ \pmu\dot\xmu -\half p^2 + i\hat\theta\dot\theta
                       +i\hat\phi\dot\phi +\hat c\dot c +\hat\kappa\dot\kappa
                       +i\hat v\dot v + i\hat\rho\dot\rho
                       +\hat\zeta\dot\zeta \Big],
\eqn\quanty$$
where
$\hat\theta = d - \psh\theta - 4\hat c\kappa$.
As the
quantum action defines a free field theory, it is easy to quantize by
imposing canonical commutation relations on the operators corresponding
to the variables
$(\pmu ,\xmu ,\hat\theta ,\theta ,\hat\phi ,\phi ,\hat c ,c ,
\hat\kappa ,\kappa ,\hat v ,v ,\hat\rho ,\rho ,\hat\zeta ,\zeta )$.
It proves useful to choose a Fock space representation for the ghost
and define a ghost vaccum $|0>$ which is annihilated by each of the
antighosts
$(\hat\kappa |0> = 0,\hat c |0> = 0,\hat v |0> = 0,\hat\rho |0> =0,
  \hat\zeta |0> = 0 )$.
It also proves useful to define a {\it twisted} ghost vacuum $|0>_g$,
where for each ghost $g$ in the subscript, that ghost is an annihilation
operator and the corresponding anti-ghost is a creation operator.
The physical states on both twisted and untwisted Fock space
should be the same, as they are {\it dual} representations of
the same spectrum.
It is then viewed the superspace coordinates $\xmu$ ,$\hat\thetaA$ and
$\phiA$ as hermitian coordinates while $\pmu =-i\grad/\grad\xmu$,
$\hat\thetaA = \grad/\grad\thetaA$ and $\hat\phiA =\grad/\grad\phiA$,
and consider states of the form $\Phi (x,\theta ,\phi ) M|\Omega >$
with wavefunction $\Phi$, where $M$ is some monomial constructed
from the (anti-)ghost and $|\Omega >$ is one of the ghost vacua.
It was found then that the ghost-independent state
$\Phi (x,\theta ,\phi )|0>$ gives the physical spectrum
consisting of the eight bosons and eight fermions which
form the Yang-Mills multiplet together with the zero-momentum
ground state which is a supersymmetry singlet.

I shall now describe a ten-dimensional spinor-twin type II superparticle
model with a spinor super-wavefunction satisfying
$$\eqalign{
p^2 \Psi_A &=0, \qquad \psh_{AC}D^C \Psi_B =0, \qquad \psh^{AB}\Psi_B =0, \cr
&(\gamma^{\mu\nu\rho\sigma})_A^{\ \ B} D^A\Psi_B =0,
\cr}\eqn\linear$$
which is equivalent to constraints \quadra .
The model is formulated in an extended ten-dimensinal
superspace with coordinates
$(\xmu ,\thetaA ,\phiA )$ where $\thetaA$ and $\phiA$ are
anticommuting Majorana-Weyl spinors, and to describe super Yang-Mills
we wish to impose the extra constraints
$d^A (\Gamma^{\mu\nu\rho\sigma})^B_{\ A} =0$ and $\hat\phi\hat\phi =0$
which can be done
by adding appropriate lagrange multiplier terms.

The spinor-twin type II superparticle action is then given by the
sum of [\two]
$$
S_0 =\int\! d\tau \Big[ \pmu\dot\xmu +i\hat\theta\dot\theta
                         + i\hat\phi\dot\phi\Big],
\eqn\snut$$
and
$$
\eqalign{
S^{''} =\int\! d\tau \Big[ -\half e p^2 + i\psi\psh d
           + i \varphi\psh\hat\phi
         & + i\Lambda_{\mu\nu\rho\sigma}d\Gamma^{\mu\nu\rho\sigma}\hat\phi\cr
         & - i\beta(\phi\hat\phi -1) +\half\hat\phi\omega\hat\phi
\Big],\cr}\eqn\sbiprim$$
where,
as usual,
$\pmu$ is the momentum conjugate to the space-time coordinate $\xmu$,
$d^A$ is a spinor introduced so that the Grassmann coordinate $\thetaA$
has a conjugate momentum $\hat\theta^A =d^A -\psh^{AB}\thetaB$,
$\phiA$ is a new spinor coordinate and $\hat\phi_A$ is its conjugate
momentum.
The fields $e$, $\psi^A$, $\varphi_A$, $\Lambda_{\mu\nu\rho\sigma}$,
$\beta$ and $\omega^{AB}=-\omega^{BA}$ are all
Lagrange multipliers (which are
also gauge fields for corresponding local symmetries) imposing the
following constraints
$$
\eqalign{ p^2 =&0, \qquad \psh d =0, \qquad \psh\hat\phi =0,\cr
         \hat\phi_A\hat\phi_B =0, \qquad
         &\phiA\hat\phi_A - 1 =0, \qquad
         d^A(\Gamma^{\mu\nu\rho\sigma})^B_{\;\ A}\hat\phi_B =0.
\cr}.\eqn\constry$$

The action \snut -\sbiprim\ is invariant under the global space-time
supersymmetry transformations [\two]
$$\delta\theta =\epsilon ,\qquad
  \delta x^\mu =i\epsilon\Gammu\theta ,\eqn\globalspin$$
(where $\epsilon_A$ is a constant Grassmann parameter)
together with a number of local symmetries.
\footnote
\spadesuit{
The symmetries divide into two types
\REF\kinds{R. E. Kallosh, A. van Proeyen and W. Troost, Phys. Lett
{\bf B212} (1988) 428.}
[\kinds]. Symmetries of the {\it first} type are those under which
a gauge field transforms into the derivative of a gauge parameter,
while of the {\it second} type are those which involves only
gauge fields.}
These include world-line
reparameterization which, when combined with a {\it trivial} symmetry,
gives the {\cal A} transformations
$$\delta e =\dot\xi ,\qquad
  \delta x^\mu =\xi p^\mu ,\eqn\trivspin$$
the other fields
being inert. There are also two
fermionic symmetries of the first kind, $\cal B$ and $\cal B$',
with fermionic spinor
parameters $\kappa^A (\tau)$ and $\varphi_A (\tau )$ given by
$$\eqalign{\delta \psi &=\dot\kappa ,\qquad
            \delta \theta ={\psh}\kappa ,\qquad
            \delta e = 4i \dot\theta\kappa ,\cr
           &\delta x^\mu =id ({\Gammu})\kappa
                      +i\theta (\Gammu ){\psh}\kappa ,
\cr}\eqn\siegspin$$
and
$$\delta {\varphi} ={\dot\zeta} + \beta\zeta ,\qquad
  \delta \phi =\zeta \psh ,\qquad
  \delta x^\mu =i\hat\phi (\Gammu )\zeta\eqn\ferspin$$
where
$\zeta_A$ is a spinor parameter.
The
bosonic symmetries associated with the gauge fields $\beta$ and
${\omega}^{AB}$ (the {\cal C} and {\cal C}' symmetries ) are defined
by
$$\eqalign{&\delta \beta =\dot\eta ,\qquad
            \delta {\hat\phi} =\eta{\hat\phi} \;,\qquad
            \delta \phi = -\eta\phi ,\cr
            \delta \omega =-2 & \eta\omega ,\qquad
            \delta \Lamnrs =\eta\Lamnrs \;,\qquad
            \delta {\varphi} = - \eta{\varphi} \;,
\cr}\eqn\etaspin$$
and
$$\delta \omega ={\dot\Upsilon} + 2\beta\Upsilon ,\qquad
  \delta \phi = i \Upsilon {\hat\phi} ,\eqn\Upsispin$$
where
$\eta$ is a bosonic parameter and $\Upsilon^{AB} =-\Upsilon^{BA}$ is a
bosonic bispinor parameter. There is also a tensor symmetry associated
with the gauge field $\Lamnrs$ (referred to as {\cal F} symmetry) with
bosonic parameter $\Sigmnrs$ and given by
$$\eqalign{&\delta \Lamnrs =\dot\Sigmnrs +\beta\Sigmnrs ,\qquad
            \delta d = -2{\hat\phi} {\Sigmash}\psh ,\cr
           &\delta \theta = -{\hat\phi} {\Sigmash} \;,\qquad
            \delta \phi = d {\Sigmash} ,\qquad
            \delta e = 4i{\hat\phi} {\Sigmash}\psi ,\cr
    &\delta x^\mu = i{\hat\phi} {\Sigmash} (\Gammu )\theta \;,
      \qquad \delta \omega = 4i{\Sigmash}\psh{\lmdash} .
\cr}\eqn\bosten$$

The gauge algebra of the symmetries {\cal A}, {\cal B}, {\cal B}',
{\cal C}, {\cal C}' and {\cal F} closes on shell. In this situation,
the Batalin and Vilkovisky procedure can be used in order to determine
the quantum action and the BRST charge. The approach which is in
principle the most straightforward, but turns out to be technically
the most complicated, involves introducing ghost fields corresponding
to each of the symmetries {\cal A}, {\cal B}, {\cal B}',
{\cal C}, {\cal C}' and {\cal F}.
The minimal set of fields that enter the BV quantisation scheme is
determined by the classical gauge symmetries, together with the
requirement that the BRST transformations of the classical fields
and the ghost should be on-shell nilpotent. This procedure fixed
much of the structure of the master action.
We introduce ghosts $(c,\kappa_1 , \zeta_1 ,\eta_1 ,v_1 ,\Sigmsh_1 )$
corresponding to the classical symmetries
\trivspin -\bosten . These ghost fields have
opposite Grassmann parity to the set of classical gauge parameters
$(\xi ,\kappa ,\varphi ,\beta ,\omega ,\Sigma )$.
The BRST transformations are
$$\eqalign {
  & s e =\dot c + 4i{\dot\theta} \kappa_1
               +  4i{\hat\phi} {\Sigmash_1}\psi \;,\qquad
    s \psi =\dot\kappa_1 \;,\qquad
    s \beta =\dot\eta_1 ,\cr
  & s x^\mu = c p^\mu + id ({\Gammu})\kappa_1
           + i\theta (\Gammu ){\psh}\kappa_1
           + i\hat\phi (\Gammu )\zeta_{1}
           + i{\hat\phi} {\Sigmash_1} (\Gammu )\theta \;\cr
  & s \theta ={\psh}\kappa_1 -\hat\phi\Sigmash_1 \;,\qquad
    s {\varphi} ={\dot\zeta}_{1} +\beta\zeta_{1} -\eta_1{\varphi} \;,\qquad
    s {\hat\phi} =\eta_1 {\hat\phi} \;,\cr
  & s \phi =\zeta_{1} \psh - \eta_1\phi + i v_1 {\hat\phi}
           + d {\Sigmash_{1}} \;,\qquad
    s \omega = -2 \eta_1\omega + {\dot v}_1 + 2\beta v_1
              + 4i{\Sigmash_1}\psh{\lmdash} \;,\cr
  & s \Lamnrs =\eta_1\Lamnrs + \dot\Sigma_{(1)\mu\nu\rho\sigma}
                             +\beta\Sigma_{(1)\mu\nu\rho\sigma} \;,\qquad
    s d = -2{\hat\phi} {\Sigmash_1}\psh ,\cr
  & s \zeta_{1} = -\eta_1\zeta_{1}  \;,\qquad
    s v_1 = - 2 \eta_1 v_1 + 2i \Sigmsh_{(1)}\psh\Sigmsh_{(1)}  \;,\qquad
    s \kappa_n = i^n\psh \kappa_{(n+1)} \;,\cr
  & s \eta_1 = \eta_1 \eta_1 , \qquad
    s \Sigmsh_{(1)} =\Sigmsh_{(1)} \eta_1, \qquad
    s c = -2i\kappa_1\psh\kappa_1
          + 4i \kappa_1\Sigmash_{(1)} \hat\phi
\cr}\eqn\brsties$$

The minimal set of fields $\Phi^A_{min}$ consists of all classical
and ghost fields that furnish a representation of the BRST algebra,
$$
\Phi^A_{min} =\Big\{x,p,e,c;\theta ,d,\psi ,\kappa_1,\dots ,\kappa_n;
               \phi ,\hat\phi ,\varphi ,\zeta_1 ;\beta ,\eta_1 ;
               \omega ,v_1;\Lambda_{\mu\nu\rho\sigma},
               \Sigma_{(1)\mu\nu\rho\sigma} \Big\}.
\eqn\minima$$
In addition
to the above minimal set of fields, gauge fixing requires the introduction
of anti-ghosts, extra-ghosts and Nakanishi-Lautrup (NL) auxiliary fields
$$
\Phi^A_{non-min} =\Big\{\hat c,\hat\kappa_1,\dots ,\hat\kappa_n,
                   \hat\zeta_1,\hat\eta_1,\hat v_1,\hat\Sigmsh_1;
                   \kappa^m_n;\pi^m_n \pi_c,\pi_1,\dots\pi_n,\pi^m_n
                   \pi_\zeta,\pi_\eta,\pi_v,\pi_\Sigma \Big\}.
\eqn\minimu$$

For each field $\Phi^A$ the BV method requires the introduction of a
corresponding anti-field $\Phi^\star_A$ of opposite Grassmann parity.
Next, we need to find a solution $ S(\Phi^A ,\Phi^\star_A) $ to the
master equation $(S,S)=0$ subject to the boundary condition
$
  S|_{\Phi^\star_A =0} = S_0 + S''$.
Yet, care must be taken, as
the grading of the fields plays an important role.
The solution to the master equation for the minimal set of field
$\Phi^A_{min}$ is
$$
S_{min} = S_0 + S'' + S_1 + S_2 , \eqn\sminima$$
where
$ S_0 + S''$ is the classical action of the spinor-twin type II
superparticle
\snut -\sbiprim . The term linear in anti-fields, $S_1$, is
$$\eqalign {
  S_1 =\int\! d\tau &\Big[ \starxmu \big(cp^\mu + i\theta\gammu\psh\kappa_1
         +id\gammu\kappa_1 + i\hat\phi\gammu\zeta_1
         -i\hat\phi\Sigmash_1\gammu\theta \big) \cr
     & + \startheta \big(\psh\kappa_1 -\hat\phi\Sigmash_1 \big)
       - \stard \big( 2\hat\phi\Sigmash_1\psh \big)
       + \stare \big( \dot c + 4i\dot\theta\kappa_1
       + 4i \hat\phi\Sigmash_1\psi \big)     \cr
     & + \spsi \big(\dot\kappa_1 \big)  + \shatphi \big(\eta_1\hat\phi \big)
       + \sphi \big(\psh\zeta_1 + i v_1\hat\phi - \eta_1\phi
       + d\Sigmash_1 \big) \cr
     & + \svarphi \big(\dot\zeta_1 + \beta\zeta_1 -\eta\varphi \big)
       + \sLamdsh \big(\eta_1\Lamdsh + \dot\Sigmash_1 + \beta\Sigmash_1 \big)
       + \sbeta \big(\dot\eta_1 \big)      \cr
     & - \szeta_1 \big(\eta_1\zeta_1 \big) + \seta_1 \big(\eta_1\eta_1 \big)
       + \somega \big(-2\eta_1\omega + \dot v_1 + 2\beta v_1
       + 4i \Sigmash_1\psh\Lamdsh \big)   \cr
     & + \skappa_n \big( i^n \psh\kappa_{(n+1)} \big)
       + \starv_1 \big(-2\eta_1 v_1 + 2i \Sigmash_1\psh\Sigmash_1 \big)
       + \sSigmash_1 \big(\Sigmash_1\eta_1 \big)    \cr
     & + \starc \big(- 2i \kappa_1\psh\kappa_1
       + 4i \kappa_1\Sigmash_1\hat\phi \big)
\Big],
\cr}\eqn\slinear$$
while
the term quadratic in antifields, $S_2$, is
$$\eqalign {
S_2 =\int\! d\tau &\Big[ 2\stare \Big( i\startheta\kappa_2
       + \sum^{+\infty}_{n=1} i^{2n+1} \skappa_n\kappa_{n+2}
       + \starxmu (i\kappa_1\gammu\kappa_1 -\theta\gammu\kappa_2) \cr
    &  + 4 \starc\kappa_1\kappa_2  \Big)
       + 4\svarphi\starc\kappa_2\Sigmash_1
       + i\starxmu (\spsi\gammu\kappa_2
       + 2 \somega\Sigmash_1\gammu\Sigmash_1)\Big].
\cr}\eqn\squadra$$
The full
master action in then given by adding to $S_{min}$ the following action
for the non-minimal fields
$$
S_{non-min}=\int\!d\tau \Big[ {\hat c}^\star \pi_c
            +\sum^{+\infty}_{n=1}
             \sum^{+\infty}_{m=0} i^n {\hat\kappa}_n^{m\star}\pi^m_n
            +\hat\zeta_1^\star\pi_\zeta +\hat\eta_1^\star\pi_\eta
            +\hat v_1^\star\pi_v + \hat\Sigmash_1^\star\pi_\Sigma
\Big]
.\eqn\snonmin$$

The corresponding quantum action $S_Q$ is then given
by substituting $\Phi^\star_A =\grad\Psi/\grad\PhiA$. The gauge fermion
$\Psi (\PhiA )$ is implemented by imposing gauge conditions on the gauge
fields rather than on the coordinates. The simplest gauge is
$$
\Psi (\PhiA )=\int\! d\tau \Big[ \hat c (e-1) +\hat\kappa\psi
                +\hat\zeta_1\varphi + \hat\eta\beta
                +\hat\Sigmsh_1\Lamdsh +\half\hat v_1\omega \Big],
\eqn\fermionx$$
where
$\hat c$, $\hat\kappa$, $\hat\zeta_1$, $\hat\eta$, $\hat\Sigmash_1$
and $\hat v_1$ are anti-ghosts fields.
This leads to the following quadratic gauge-fixed quantum action.
$$
S_Q =\int\! d\tau \Big\{\pmu\dot\xmu + i\hat\theta\dot\theta
       +i\hat\phi\dot\phi +\half p^2 +\hat c\dot c
       +\hat\kappa_1\dot\kappa_1 +\hat\zeta_1\dot\zeta_1
       +\hat\Sigmash_1\dot\Sigmash_1 +\hat\eta_1\dot\eta_1
       +\hat v_1\dot v_1 \Big\}.
\eqn\quadryx$$
where
$\hat\theta = d-\psh\theta -4i\hat c\kappa_1$. This is invariant
under the modified BRST transformations given by
$\hat s \PhiA = \grad_l S/\grad\Phi^\star_A
                 \Big|_{\Phi^\star_A =\grad\Phi/\grad\phiA } $,
which are generated by
the following conserved $(\dot Q_{BRST}=0)$ and nilpotent
$(Q^2_{BRST} =0)$ BRST charge
$$\eqalign {
 Q_{BRST} & =\half c p^2 + 2id\psh\kappa_1 + 2i \hat\phi\psh\zeta_1
            - 2i d\hat\phi\Sigmash_1 -\hat\phi v_1\hat\phi   \cr
          & -\hat\phi\eta_1\phi + \hat\zeta_1\zeta_1\eta_1
            +\hat\Sigmash_1\Sigmash_1\eta_1 +\hat\eta_1\eta_1\eta_1
            + 2 \hat v_1 v_1\eta_1 - 2 \hat c\hat\theta\kappa_2   \cr
          & - 2i \hat c\kappa_1\psh\kappa_1 + i \hat\kappa_1\psh\kappa_2
            + 4 \hat c\hat\zeta_1\kappa_2\Sigmash_1
            + 2i \hat\kappa_1\hat c\kappa_3
            + 2i \hat v\Sigmash_1\psh\Sigmash_1 \;.
\cr}\eqn\brstcharge$$

As the action \quadryx\ is free, the model can be quantized canonically
by replacing each of the fields by an operator and imposing canonical
(anti-) commutation relations on the conjugate pairs
$(\pmu ,\xmu )$, $(\hat\theta ,\theta )$, $(\hat\phi ,\phi)$,
$(\hat c, c)$, $(\hat\kappa_1 ,\kappa_1)$, $(\hat\zeta_1 ,\zeta_1)$,
$(\hat\Sigmash_1 ,\Sigmash_1)$, $(\hat\eta_1 ,\eta_1)$ and
$(\hat v_1, v_1)$. Then a state is physical if it is annihilated
by the BRST charge. The cohomology classes can be classified according
to their total ghost number and the physical states are taken to be
the cohomology class of some definite ghost number.
We consider two distinct Fock space representations of the ghost system,
the {\it untwisted} one in which the ghost ground state $|0>$ is
annihilated by each of the antighosts
$(\hat\kappa_1 |0> = 0,\hat c |0> = 0,\hat v_1 |0> = 0,\hat\eta_1 |0> =0,
  \hat\Sigmash_1 |0>=0, \hat\zeta_1 |0> = 0 )$,
and the {\it twisted}
one in which antighosts are creation operators and ghosts are
annihilation operators.
It is viewed $x$, $\theta$ and $\phi$ as hermitian coordinates while
$\pmu =-i\grad/\grad\xmu$, $\hat\thetaA = \grad/\grad\thetaA$ and
$\hat\phiA =\grad/\grad\phiA$, and
consider states of the form $\Phi (x,\theta ,\phi )\;M\;|\Omega >$
with wavefunction $\Phi$, where $M$ is some monomial constructed
from (anti-)ghosts and $|\Omega >$ is one of the ghost ground state.
The wave function $\Phi (x,\theta ,\phi )$ satisfies
the following conditions
$$\eqalign { &
    p^2 \Phi =0, \qquad \psh d\Phi =0, \qquad \psh\hat\phi\Phi =0,  \cr
   \hat\phi\hat\phi\Phi &=0, \qquad (\phi\hat\phi -1)\Phi=0, \qquad
   d\Gamma^{\mu\nu\rho\sigma}\hat\phi\Phi =0,
\cr}\eqn\constries$$
which
are precisely the the ones discussed in [\two].
Therefore, the BRST cohomology class with no ghost dependence gives
a physical spectrum consisting of 8 bosons and 8 fermions which
form the super Yang-Mills multiplet. The monomial $M$ cohomology
classes have to be investigated in both spinor-twin type I and type II
models.

\ack {I would like to thank C.M. Hull and N. Berkovits
      for useful discussions. It is also acknowledge to "Centro
      Latinoamericano de Fisica" for kind hospitality.}

\refout

\end